\documentclass[runningheads]{llncs}

\usepackage{eccv}

\usepackage{eccvabbrv}

\usepackage{graphicx}
\usepackage{booktabs}
\usepackage{amsmath}
\usepackage{makecell}

\usepackage[accsupp]{axessibility}

\usepackage{bm}
\usepackage{xspace}
\usepackage{bbding}

\usepackage{caption}
\usepackage{subcaption}
\usepackage{wrapfig}
\usepackage{mathtools}
\usepackage{ifthen}
\usepackage[most]{tcolorbox}
\usepackage[export]{adjustbox}

\usepackage{hyperref}

\usepackage{orcidlink}
\renewcommand{\orcidID}[1]{\orcidlink{#1}}

\begin{document}

\newcommand{\revise}[1]{#1}

\newcommand{\website}[1]{{\tt #1}}
\newcommand{\program}[1]{{\tt #1}}
\newcommand{\benchmark}[1]{{\it #1}}

\newcommand*\circled[2]{\protect\tikz[baseline=(char.base)]{
            \protect\node[shape=circle,fill=black,inner sep=1pt] (char) {\textcolor{#1}{{\footnotesize #2}}};}}

\newcommand{\figline}{{\vspace*{.05in}\hline}}

\newcommand{\Sect}[1]{Sec.~\ref{#1}}
\newcommand{\Fig}[1]{Fig.~\ref{#1}}
\newcommand{\Tbl}[1]{Tbl.~\ref{#1}}
\newcommand{\Eqn}[1]{Eqn.~\ref{#1}}
\newcommand{\Apx}[1]{Apdx.~\ref{#1}}
\newcommand{\Alg}[1]{Algo.~\ref{#1}}

\newcommand{\specialcell}[2][c]{\begin{tabular}[#1]{@{}c@{}}#2\end{tabular}}
\renewcommand{\note}[1]{}

\newcommand{\proj}{\textsc{ControlHair}\xspace}
\newcommand{\algo}{\textsc{SpaRW}\xspace}
\newcommand{\mode}[1]{\underline{\textsc{#1}}\xspace}
\newcommand{\sys}[1]{\underline{\textsc{#1}}}

\newcommand{\no}[1]{}
\newcommand{\RNum}[1]{\uppercase\expandafter{\romannumeral #1\relax}}

\newcommand{\cg}{\mathcal{G}}
\newcommand{\cp}{\mathcal{P}}
\newcommand{\cs}{\mathcal{S}}
\newcommand{\crr}{\mathcal{R}}
\newcommand{\cL}{\mathcal{L}}
\newcommand{\cM}{\mathcal{M}}
\newcommand{\cF}{\mathcal{F}}

\newcommand{\bs}{\mathbf{s}}
\newcommand{\be}{\mathbf{e}}
\newcommand{\bi}{\mathbf{I}}
\newcommand{\bj}{\mathbf{J}}
\newcommand{\bsi}{\mathbf{SI}}

\newcommand{\degree}[1]{#1\textdegree\xspace}
\newcommand{\corrauth}{\textsuperscript{\raisebox{0.15ex}{\tiny\Envelope}}}

\graphicspath{{figs/}{supp/figs/}}

\title{\proj: Synergizing Physics Simulator and Video Diffusion for Controllable Dynamic Hair Rendering}

\titlerunning{ControlHair}


\author{Weikai Lin\inst{1}\thanks{This work was primarily conducted during an internship at Pixocial Technology.}\corrauth\,\orcidID{0000-0003-3537-4857} \and
Haoxiang Li\inst{2}\,\orcidID{0009-0006-1525-3942} \and
Yuhao Zhu\inst{1}\,\orcidID{0000-0002-2802-0578}}

\authorrunning{W. Lin et al.}

\institute{$^{1}$University of Rochester{, USA} \quad $^{2}$Pixocial Technology{, USA}\\
\email{wlin33@ur.rochester.edu, haoxiang@pixocial.com, yzhu@rochester.edu} \qquad \corrauth Corresponding author.}

\maketitle

\begin{abstract}
Hair simulation and rendering are challenging due to complex strand dynamics, diverse material properties, and intricate light–hair interactions.
Recent video diffusion models can generate high-quality videos, but they lack fine-grained control over hair dynamics.
We present \proj, a hybrid framework that integrates a physics simulator with conditional video diffusion to enable precise and controllable dynamic hair rendering.
\proj adopts a three-stage pipeline: it first encodes physics conditions into per-frame geometry using a simulator, then extracts per-frame control signals, and finally feeds control signals into a video diffusion model to generate videos with desired hair dynamics.
This cascaded design decouples physics reasoning from video generation, supports diverse physics, and makes training the video diffusion model easy.
Trained on a curated 10K video dataset, \proj outperforms text- and pose-conditioned baselines, delivering precisely controlled hair dynamics.
We also demonstrate three use cases of \proj, including dynamic hairstyle try-on, bullet-time effects, and cinemagraphic.
Project page: \href{https://linwk20.github.io/controlhair-web}{https://linwk20.github.io/controlhair-web}.


\end{abstract}

\section{Introduction}

Human hair simulation and rendering are long-standing research problems in computer graphics~\cite{petrovic2005volumetric, bergou2008discrete, hsu2023sag, wang2023neuwigs, luo2024gaussianhair}, with applications ranging from films and games~\cite{iben2013artistic, EpicGroom2020} to VR/AR and digital humans~\cite{xu2024gaussian,ward2006simulation}.
However, hair is exceptionally complex, involving tens of thousands of interacting strands as well as intricate material properties and light scattering. 
This complexity makes simulation and rendering challenging and computationally demanding.

Video diffusion models~\cite{wan2025wan, yang2024cogvideox, openai_sora, kong2024hunyuanvideo, hacohen2024ltx} have demonstrated the ability to generate high-quality videos rivaling professional production. 
Beyond common text- and image-conditioned generation, recent works have explored fine-grained and multi-modal controls to enable practical usecases. 
Examples include conditioning video generation with human pose~\cite{wang2024unianimate,ma2024follow, wang2025unianimate}, lighting~\cite{ye2025stylemaster}, or speech and music~\cite{chen2025echomimic,tian2025emo2}.
This suggests a promising direction for dynamic hair rendering: one can formulate it as a conditioned generation problem, where physics parameters serve as input conditions and the output is a video with desired hair dynamics.

\begin{figure*}[t]
    \centering
    \includegraphics[width=1\textwidth]{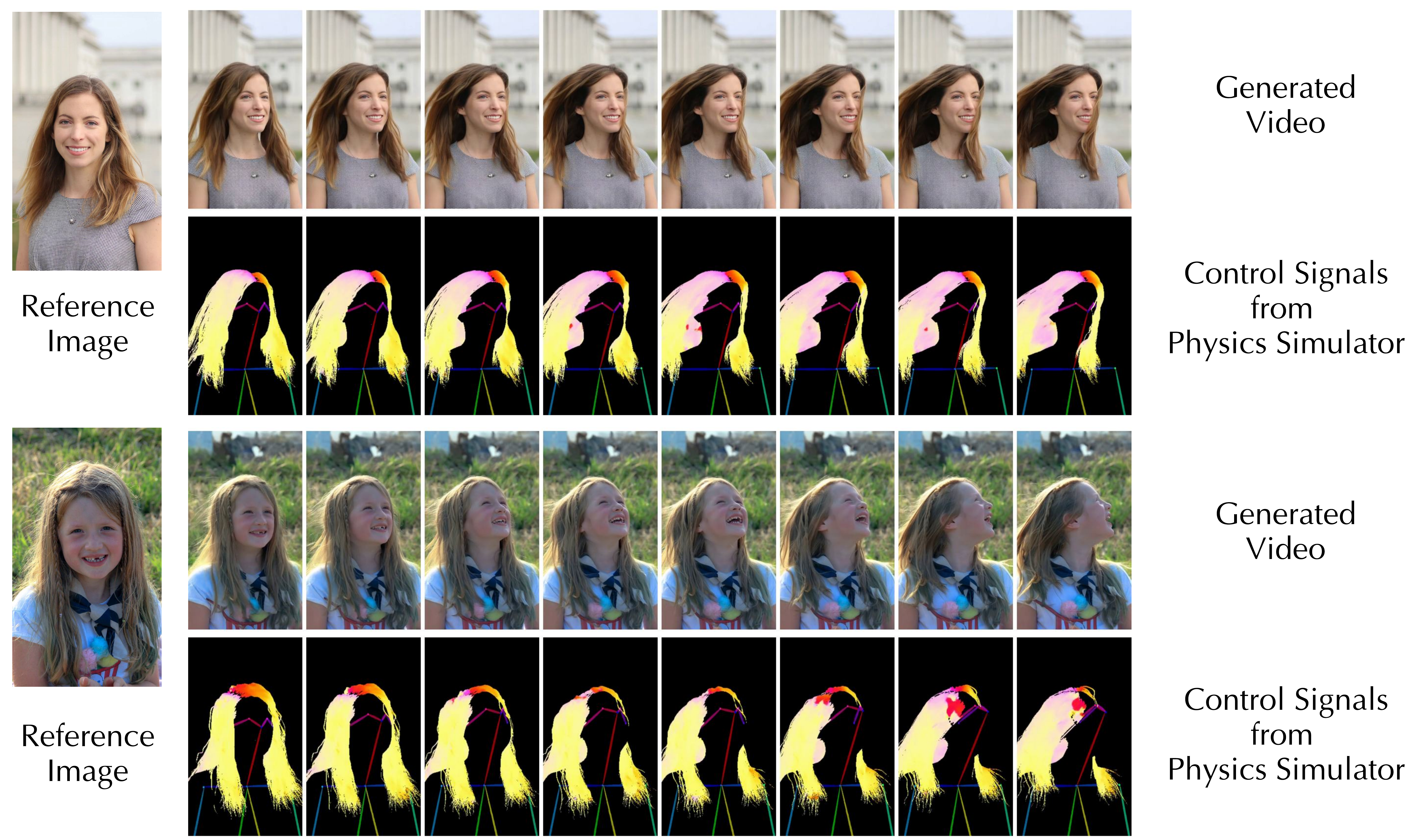}
    \caption{Given a reference image, \proj converts physics conditions (e.g., hair stiffness, wind, or human motion) into per-frame control signals using a physics simulator.
  These signals, combined with a reference image, guide a diffusion model to generate photorealistic videos with controlled hair dynamics.
      }
  \label{fig:teaser}
  \end{figure*}

{\Fig{fig:teaser} illustrates the overall setting: user-specified physics conditions are converted into geometry controls that guide video generation from a reference image.}
However, none of the existing conditional video diffusion methods support fine-grained control over dynamic hair rendering.
This limitation arises from two challenges:
(1) There are numerous physics parameters that influence hair dynamics, including intrinsic hair properties (e.g., mass, stiffness, damping) and external forces (e.g., gravity, wind). There is still no general encoding mechanism that can accommodate the diverse conditions.
(2) training such conditional models would require videos annotated with ground-truth physics. 
However, physics like hair property or wind direction/strength are  difficult to infer  from videos.

To overcome the above challenges and enable dynamic hair rendering using diffusion models, we first formulate hair simulation and rendering as an image animation problem in the context of conditional video generation (\Sect{sec:problem_formulation}). 
We then introduce \proj, a video diffusion framework for physically-based hair rendering, targeting portrait and digital-human hair rendering.

\proj uses a physics simulator to convert physical parameters into per-frame geometry, which then conditions the diffusion model. This design makes \proj agnostic to the specific physics input and reduces annotation cost, since training only requires per-frame geometry rather than hard-to-infer physical parameters. \proj consists of three key components:

\textbf{Simulator-based Physics Encoding.}
Physics conditions for hair rendering span a wide range of intrinsic parameters and external forces, often specific to the scene, task, or physics model.
A straightforward approach is to directly feed these parameters into a diffusion model. 
However, this introduces two key challenges. 
First, there may be tens of parameters, making it difficult for the diffusion model to learn 
the effects of different physics factors.  
Second, any change in input physics requires redesigning the diffusion model, recollecting training data, and retraining the model, which  limits flexibility and generality.

\proj addresses above issues by cascading a physics simulator before the diffusion model.
We first estimate a 3D hair model from the input image.  
The simulator then takes physics parameters as input and use them to generate per-frame hair geometry (\Sect{sec:method:sim}).
This simulation+diffusion hybrid approach has two advantages.  
First, it offloads dynamics generation to the simulator, relaxing the burden of physics reasoning in the diffusion model.  
Second, we unify the diffusion model input as per-frame geometry. 
This decouples the diffusion model from specific physics settings: whenever the physics conditions change, we only need to swap the simulator, without redesigning downstream modules.

\textbf{Per-frame Control Signals Extraction.}
After obtaining the per-frame hair geometry, we convert it into control signals for the video diffusion model (\Sect{sec:method:ctrl_sig}). 
We first project the 3D geometry onto 2D images along the desired camera trajectory.  
From these projections, we extract per-frame hair strand maps~\cite{zheng2023hairstep} and human poses~\cite{yang2023effective} as control signals, which guide the video diffusion model on a per-frame basis.
This control-signal design is motivated by three reasons. 
First, by projecting 3D geometry to 2D, we can control the poses of the generated frames.  
Second, strand maps and human poses depend only on geometry, which is available from simulation.  
Third, these per-frame control signals can be easily extracted from 2D frames in real-world videos, making data annotation easy.
This greatly reduces the difficulty of training data annotation compared to inferring precise 3D geometry or underlying physics.

\textbf{Conditioned Video Diffusion for Hair Rendering.}
Finally, we design a conditioned video diffusion model that takes a reference image (for appearance) and the per-frame control signals to generate desired RGB video (\Sect{sec:method:diffusion}). 
The reference image is encoded using Wan encoder~\cite{wan2025wan}, while the per-frame control signals are encoded with a  3D convolution neural network. 
For training, we collect a large set of high-quality real-world videos containing hair and annotate them with strand maps and human poses to construct input control signals. 
The reference image is sampled from the video.  
The model is then trained to map the reference image and control signals to RGB videos.
To leverage the prior of pretrained models, our model is finetuned from pretrained model using LoRA~\cite{hu2021lora}, which adapts the model with low-rank updates to the weight matrices.

Unlike purely data-driven models that lack fine-grained control over the physics in generated videos, \proj conditions on user-defined physics simulation, enabling detailed, interpretable, and editable dynamics.
Experiments show that \proj outperforms existing video diffusion models in controllable dynamic hair rendering.
We further showcase its applicability with diverse use cases, including dynamic hair try-on, bullet-time effects, and cinemagraphic animation.
Our contributions are summarized as follows:

\begin{itemize}
    \item We formulate controllable dynamic hair rendering as a conditional video generation problem with physics inputs.
    \item We introduce the first video diffusion framework that incorporates a physics simulator to support  fine-grained hair-related physics control.
    \item We design a control signals extraction pipeline to bridge simulators and diffusion models, enabling pose control and simplifying training data annotation.
    \item We design a conditional video diffusion model for hair rendering and construct a large-scale, high-quality  training set.
\end{itemize}

\section{Related Work}

\subsection{Controllable Video Diffusion Models}
Video diffusion has achieved significant quality improvements through both data and computational scaling, with powerful closed-source models~\cite{openai_sora, germanidis2024gen3, polyak2024movie, deepmind2025veo} and open-source counterparts~\cite{kong2024hunyuanvideo, yang2024cogvideox, zheng2024open, wan2025wan}.
Among them, the Wan series~\cite{wan2025wan} has achieved extraordinary quality.
Our diffusion model builds on the Wan~2.1 architecture, with modified interfaces and extra encoder modules to enable controllable dynamic hair rendering.  
Beyond text and image conditions, recent works incorporate additional control signals such as human pose~\cite{wang2024unianimate, wang2025unianimate}, sketch~\cite{xing2024tooncrafter,jiang2025vidsketch}, camera trajectory~\cite{he2024cameractrl}, audio~\cite{chen2025echomimic,tian2025emo2}, style~\cite{ye2025stylemaster}, and lighting~\cite{liang2025diffusion}.
PhysAnimator~\cite{xie2025physanimator} uses a simulator to guide dynamics generation, but is limited to 2D mesh deformation.
None of the existing works investigate control over the \emph{physical dynamics} of fine human hair structures.
Our work uniquely combines video diffusion with a physics simulator to encode hair-related physics.

\subsection{Hair Simulation and Rendering}
Human hair simulation and rendering are long-standing challenges in computer graphics. 
Simulation must account for tens of thousands of interacting strands, while rendering requires handling complex materials and light-scattering effects.

Various physics simulation methods have been proposed for hair dynamics, such as mass-spring chains~\cite{rosenblum1991simulating}, rigid multi-body chains~\cite{anjyo1992simple}, discrete elastic rods~\cite{bergou2008discrete}, Material Point Method~\cite{fei2021principles}, hybrid approaches~\cite{hsu2023sag}, and learning-based methods~\cite{lin2025neuralocks}.
These simulations are built on different physics models tailored for different tasks. 
Ideally, we should select the most suitable model given a task. 
However, the diversity of physics parameters across models makes it difficult to design a single diffusion model for all possible physics parameters input.
For hair rendering, prior works have explored physically based~\cite{petrovic2005volumetric,marschner2003light,moon2008efficient,zinke2008dual} and data-driven methods~\cite{wang2023neuwigs,luo2024gaussianhair}.
However, they either require complex modeling or fail to produce photorealistic results.

Our framework is compatible with arbitrary physics parameters and simulators because we unify the diffusion model input as per-frame  geometry. 
This allows us to switch freely across different simulation models for a given task without redesigning the diffusion model.
For hair rendering, we adopt video diffusion priors to avoid complex hair material and light modeling, while producing photorealistic and temporally consistent results.

Related hair-avatar systems, such as HHAvatar~\cite{liao2025hhavatar}, DGH~\cite{wang2026dgh}, and HAAR~\cite{sklyarova2024text}, focus on reconstructing or generating hairstyles, and are complementary to \proj's focus on controllable dynamic hair rendering.

\section{Problem Formulation}
\label{sec:problem_formulation}

We first formulate dynamic hair rendering within a conditional video generation framework.  
This formulation extends naturally to physics-conditioned generation of other deformable objects.  
Concretely, we model it as image animation problem:the inputs include a reference image $I_{\text{ref}}$, physics parameters $\mathcal{P}$, and a camera trajectory $\mathcal{T}_{\text{cam}}$. 
The output is a video $\text{V}$ that preserves the reference appearance while following the dynamics dictated by $\mathcal{P}$ and camera trajectory defined by $\mathcal{T}_{\text{cam}}$, which can be written as:
\begin{equation}
\text{V} = \mathcal{G}(\text{I}_{\text{ref}}, \mathcal{P}, \mathcal{T}_{\text{cam}})
\label{eq:form1}
\end{equation}
where $\mathcal{G}$ denotes the conditional video generator.
\begin{table}[t]
\centering
\caption{Input conditions for conditional video generation.}
\label{tbl:conditions}
\resizebox{\textwidth}{!}{%
\begin{tabular}{llll}
\toprule
\textbf{Reference Image} & \textbf{Hair Properties} & \textbf{External Forces} & \textbf{Camera Trajectory} \\
\midrule
\makecell[l]{Human, hair,\\ environment lighting} & \makecell[l]{Mass, stiffness,\\ damping, thickness} & \makecell[l]{Wind, gravity,\\ friction, human motion} & \makecell[l]{Pose per frame,\\ viewpoint changes} \\
\bottomrule
\end{tabular}}
\end{table}
We now specify the physics parameters $\mathcal{P}$ that govern hair dynamics.  
We categorize them into two groups: intrinsic hair properties ($\mathcal{P}_{\text{hair}}$) and external forces ($\mathcal{F}$).  
The former characterizes hair geometry and physical attributes that determine how strand responses to external stimuli (e.g. mass, stiffness).  
The latter captures external forces in the scene such as wind or gravity.
We summarize the conditions and examples in \Tbl{tbl:conditions}. 
Note that the physics parameters $\mathcal{P}$ are not fixed, as they depend on the task and the underlying physical modeling principles.  
We can rewrite the formulation in \Eqn{eq:form1} as:
\begin{equation}
\text{V} = \mathcal{G}\big(\text{I}_{\text{ref}},\, \mathcal{P}_\text{hair},\, \mathcal{F}, \mathcal{T}_{\text{cam}}\big)
\label{eq:form2}
\end{equation}
where  $\mathcal{G}$ must fuse $I_{\text{ref}}$, $\mathcal{P}_{\text{hair}}$, $\mathcal{F}$, and $\mathcal{T}_{\text{cam}}$ to produce a video in which appearance, hair dynamics, and camera motion are faithfully controlled.  
However, directly feeding these conditions into $\mathcal{G}$ makes generation difficult.  
To address this, people usually employ some form of encoder that transforms the conditions into control signals better suited for $\mathcal{G}$.
For instance, one of the most widely adopted approaches, ControlNet~\cite{zhang2023adding}, unifies all conditions into control signals that share the same shape as the target output, i.e., the video $\text{V}$ in our formulation.
Based on this idea, we can rewrite \Eqn{eq:form2} as:
\begin{equation}
\text{V} = \mathcal{G}\big(E_I(\text{I}_{\text{ref}}),\, E_{\text{hair}}(\mathcal{P}_{\text{hair}}),\, E_F(\mathcal{F}),\, E_T(\mathcal{T}_{\text{cam}})\big)
\label{eq:form3}
\end{equation}
where $E_I$, $E_{\text{hair}}$, $E_F$, $E_T$ denote encoders that transform the corresponding input conditions into control signals consumed by $\mathcal{G}$.
Prior work has studied encoding the reference image for appearance control~\cite{wan2025wan} and encoding camera trajectories for pose control~\cite{he2024cameractrl}.  
However, accurately encoding dynamics-related physics, such as $\mathcal{P}_{\text{hair}}$ and $\mathcal{F}$, remains unexplored.
The difficulty arises because of two reasons: (1) the effects of physics parameters are often highly entangled, making it difficult for a model to disentangle their effects.  
For example, lower damping, weaker gravity, or stronger wind can all lead to similar visual outcomes, such as elevated hair strands.  
(2) The parameter set $\mathcal{P}$ may vary across physics models designed for different tasks, which makes it challenging to design a single diffusion model that handles all inputs.
Based on these reasons, an ideal condition encoding mechanism must both ease model learning and remain flexible to accommodate diverse physics inputs.

\begin{figure*}[t]
    \centering
    \includegraphics[width=1\textwidth]{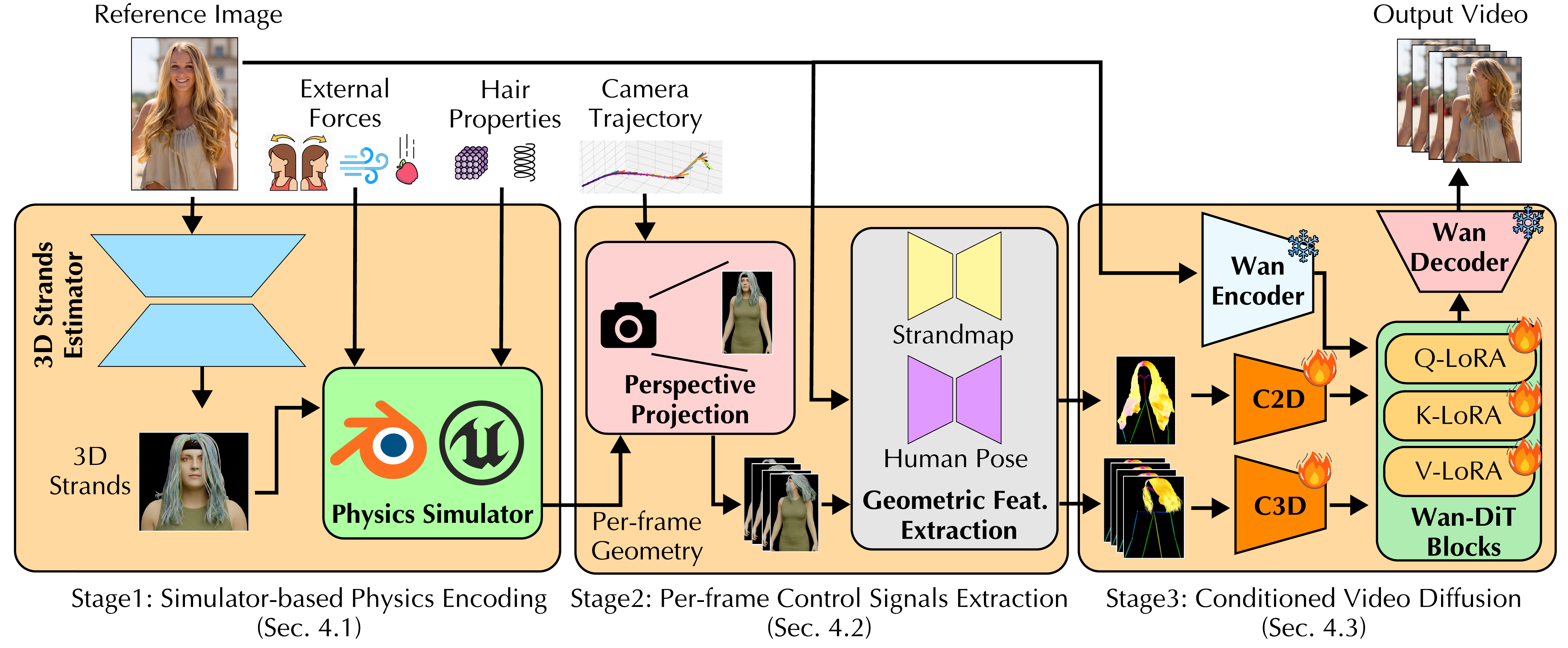}
    \caption{
  \proj enables controllable dynamic hair rendering via three stages:
  (1) Simulator-based Physics Encoding (\Sect{sec:method:sim}),
  (2) Per-frame Control Signal Extraction (\Sect{sec:method:ctrl_sig}), and
  (3) Conditioned Video Diffusion for Hair Rendering (\Sect{sec:method:diffusion}).
  }
    \label{fig:overview}
  \end{figure*}
\section{Method}
\label{sec:method}
In \proj, we investigate  physics encodings that is both general and precise, which also reduce training data annotation cost.
\proj adopts a three-stage pipeline as shown in \Fig{fig:overview}.  
First, we encode physics conditions into per-frame hair geometry using a physics simulator (\Sect{sec:method:sim}).  
Second, we extract per-frame control signals by (1) projecting the geometry onto the 2D image along the camera trajectory and (2) extracting strand maps and human poses (\Sect{sec:method:ctrl_sig}).  
Finally, the per-frame control signals, together with the reference frame, are fed into a  video diffusion model to generate the output video (\Sect{sec:method:diffusion}).  

We can also map stages in \proj to \Eqn{eq:form3}.  
The first two stages encode physics conditions and camera trajectories to produce control signals, corresponding to $E_{\text{hair}}, E_F, E_T$.
The final stage employs $E_I$ from Wan~2.1~\cite{wan2025wan} to encode the reference image and uses a video diffusion model as $\mathcal{G}$ that jointly processes all encoded control signals.
We describe each stage in detail below.

\subsection{Simulator-based Universal Physics Encoding}
\label{sec:method:sim}

\textbf{Motivation.} 
Encoding physics into video generation is challenging.  
First, many parameters have highly entangled effects. 
For example, decreasing hair mass, increasing wind force, or reducing gravity can all amplify hair motion, making it difficult to disentangle their contributions and learn the correct physics.  
Second, different target effects often require different physical modelings, which introduce distinct physics parameters. 
It is therefore difficult to design a universal video diffusion  that can handle arbitrary parameters across diverse physics models.

To address the above challenges, \proj cascades a physics simulator before running video diffusion (Stage 1 in \Fig{fig:overview}).
The simulator takes as input user-defined physics parameters, i.e., intrinsic hair properties and external forces (\Tbl{tbl:conditions}), together with a 3D hair strand model estimated from the reference image.
It then performs time-step simulations on the 3D hair model to infer per-frame hair geometry.  
The process can be formulated as:
\begin{equation}
    \text{H}_{1:\text{T}} = \mathcal{S}(\text{H}, \mathcal{P}_{\text{hair}}, \mathcal{F}),
\end{equation}
where $\text{H}$ is the 3D hair model estimated from the reference image,  
$ \mathcal{P}_{\text{hair}}$, $\mathcal{F}$ represent hair intrinsic properties and external forces, respectively,  
and $\mathcal{S}$ is a physics simulator 
$\text{H}_{1:\text{T}}$ is per-frame hair geometry over T frames.

We encode $\mathcal{P}_{\text{hair}}$ by configuring hair-related physical parameters in the simulator.  
For external forces $\mathcal{F}$, standard settings cover most cases, such as wind strength/direction and gravity.  
To account for the effect of human motion (a component of $\mathcal{F}$) on hair dynamics, we instantiate a digital human in the simulator, attach hair strands to its scalp, and drive the character with the desired motions.  
This setup allows human motion to directly influence hair behavior.
The simulator finally outputs $\text{H}_{1:\text{T}}$.  
The $\text{H}_{1:\text{T}}$ is then encoded into per-frame control signals (\Sect{sec:method:ctrl_sig}), which will be used to guide the video generation (\Sect{sec:method:diffusion}).
Note that we are free to switch among different simulators ($\mathcal{S}$) and physics conditions ($\mathcal{P}_{\text{hair}}$, $\mathcal{F}$) as long as the output of the simulator is per-frame hair geometry.

This cascaded, decoupled design offers two key advantages.  
First, it delegates the physics reasoning to simulator, leaving the video diffusion model to focus on photorealistic inpainting based on simulated geometry, thereby reducing its burden.  
Second, it unifies diverse physics parameters into a common representation: per-frame hair geometry.  
As a result, users can modify the parameter set for different tasks simply by switching to a suitable simulator, without altering the downstream design.
However, estimating 3D hair strands from a single image may introduce geometric errors, as it is an under-constrained problem.  
These errors could propagate into the generated video.
Fortunately, we observe that the diffusion model, guided by the reference image, often corrects mismatches caused by this imperfect reconstruction.  
We further discuss this in \Sect{sec:discussion}.

\subsection{Per-frame Control Signals Extraction}
\label{sec:method:ctrl_sig}

After obtaining per-frame hair geometry $\text{H}_{1:\text{T}}$ from the simulator, we convert it into control signals for video diffusion.
Here, the inputs include $\text{H}_{1:\text{T}}$ and a human model sequence $\text{M}_{1:\text{T}}$, defined by the sequence of human  poses in $\mathcal{F}$.
This stage then outputs per-frame control signals for the video diffusion model, which control both hair dynamics and human motion.
Inspired by recent controllable generation~\cite{zhang2023adding, wang2024unianimate, wang2025unianimate}, we define the control signals to have the same shape as the output video ($\text{T} \times \text{H} \times \text{W}$), thereby providing pixel- and frame-wise control.

We impose two requirements on the control signals: (1) they must reflect the desired camera trajectory $\mathcal{T}_{\text{cam}}$ to allow camera pose control, and (2) they should depend only on geometry, as the physics simulator produces purely geometric results.
\proj generates the control signals in two steps (Stage~2 in \Fig{fig:overview}):  
(1) trajectory-aware 3D-to-2D perspective projection and  
(2) geometric feature extraction.    
The first step ensures that the camera trajectory is encoded, and the second ensures the control signals depend only on geometry.

\textbf{Trajectory-aware perspective projection.}
To generate control signal for frame $\text{t}$, we first project the hair geometry and  human model at time $\text{t}$ to 2D image using the  camera pose $\text{T}_{\text{cam}, \text{t}}$, producing a projected image $\text{I}_{\text{proj}, \text{t}}$.
Formally:
\begin{equation}
    \text{I}_{\text{proj}, \text{t}} = \Pi(\text{H}_\text{t}, \text{M}_\text{t}; \text{T}_{\text{cam}, \text{t}})
\end{equation}
where $\Pi(\cdot)$ denotes the perspective projection, which projects 3D objects onto a 2D image with the same resolution as the target video.
$\text{H}_\text{t}$ is the simulated hair geometry at time $\text{t}$, and $\text{M}_\text{t}$ represents the human model at time $\text{t}$, decided by the human pose at time $\text{t}$. 
This operation is applied to all frames along the camera trajectory, yielding $\text{I}_{\text{proj},1:\text{T}}$.
This design enables explicit control over camera pose, as validated by the bullet-time effect in \Fig{fig:bullet}.

\begin{figure*}[t]
  \centering
  \begin{minipage}[t]{0.48\textwidth}
    \centering
    \includegraphics[width=\textwidth]{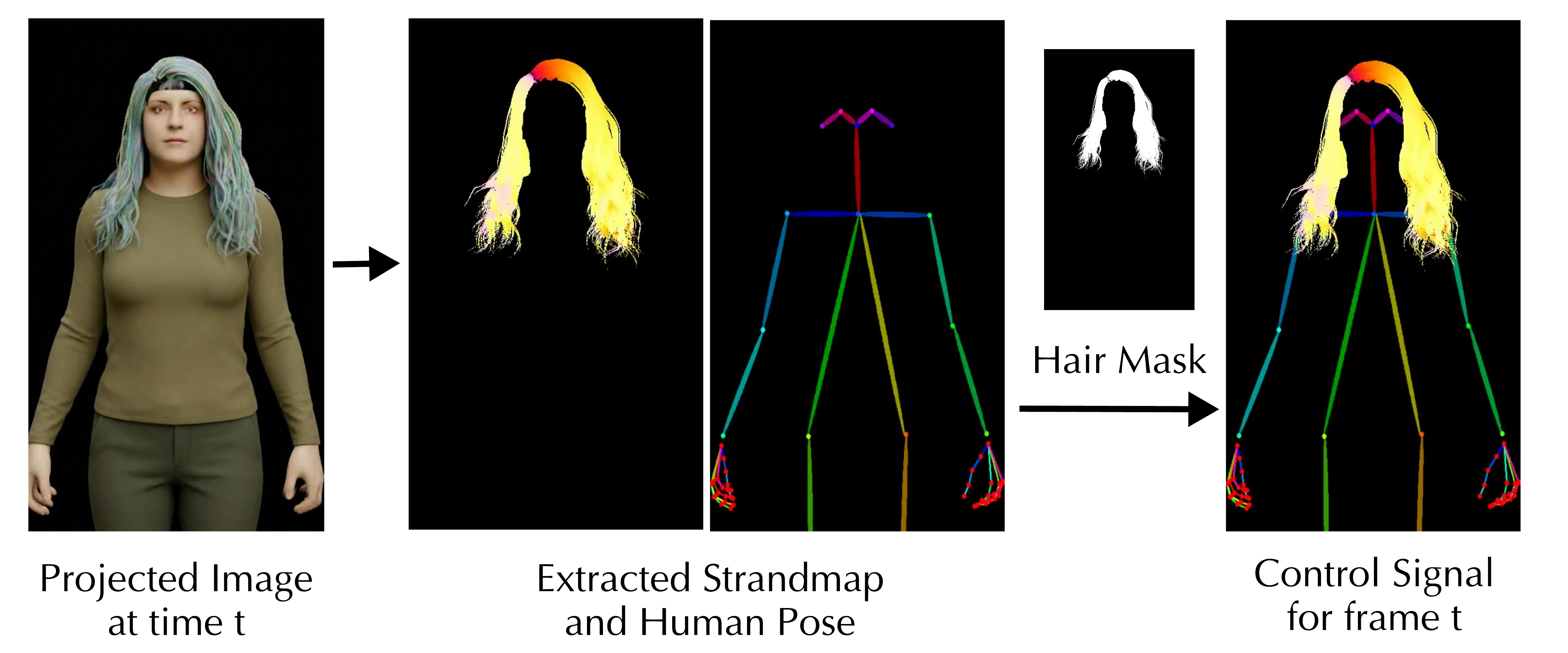}
    \caption{Example of extracting control signals from a projected image.
    Control signals are then fed into the diffusion model.}
    \label{fig:cs}
  \end{minipage}
  \hfill
  \begin{minipage}[t]{0.48\textwidth}
    \centering
    \includegraphics[width=\textwidth]{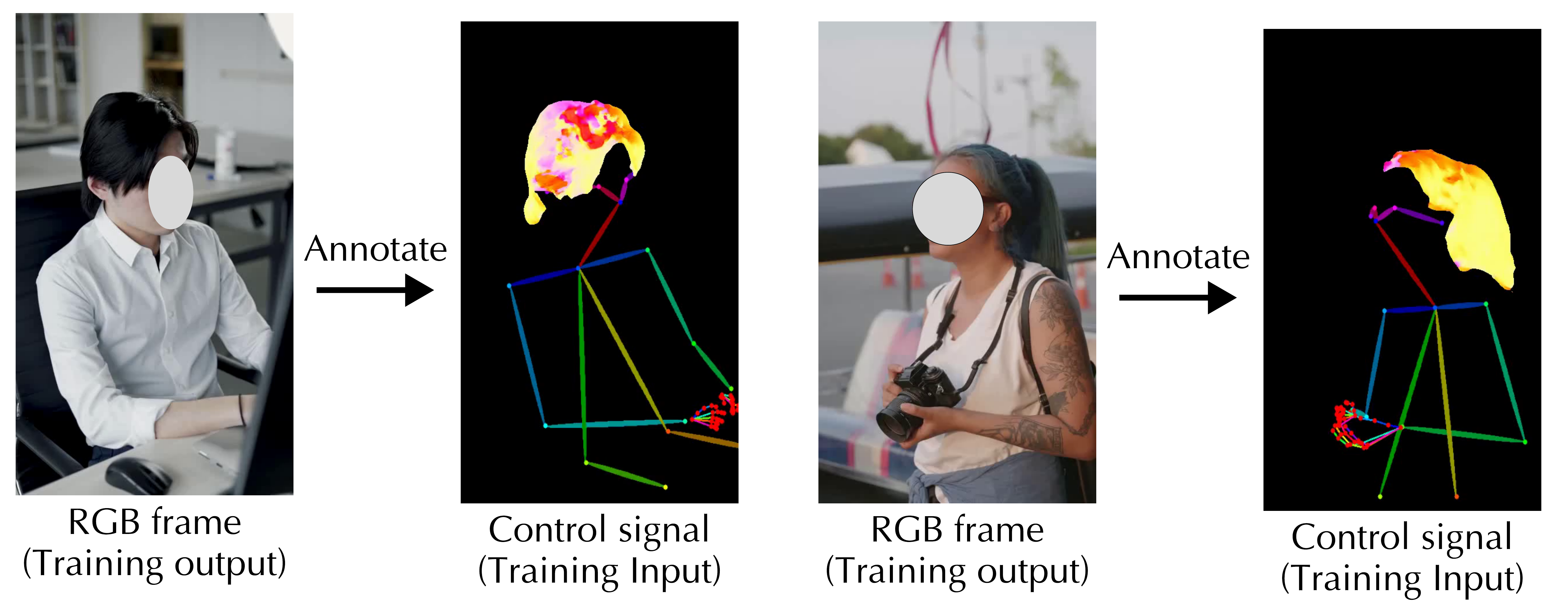}
    \caption{Training data annotation. Extracting our control signals from  videos is easier than inferring the physics.}
    \label{fig:annotation}
  \end{minipage}
\end{figure*}

\textbf{Geometric feature extraction.}
After obtaining  $\text{I}_{\text{proj},1:\text{T}}$, we extract geometric features as control signals. 
For hair, we use strand map~\cite{zheng2023hairstep}, which resemble orientation map but disambiguate directional conflicts.  
For human pose, we use sparse 2D Dwpose encodings~\cite{yang2023effective}.  
Since the simulated 3D geometry is available at inference time, strand maps can also be extracted analytically, which may be more efficient. 
However, since analytical strand maps are unavailable for the training data, we use estimated strand maps as in HairStep~\cite{zheng2023hairstep} and apply the same estimator at inference time to maintain training-inference consistency.

The two features are then merged into a three-channel RGB control image using a hair mask $B_{\text{hair}}$ that indicates where hair occludes the human body.
Formally, the geometric extraction $G$ can be written as:
\begin{equation}
    \text{C}_{\text{t}} = G(\text{I}_{\text{proj},\text{t}})
    = \Phi_{\text{strand}}(\text{I}_{\text{proj},\text{t}}) \cdot B_{\text{hair}} 
    + \Phi_{\text{pose}}(\text{I}_{\text{proj},\text{t}}) \cdot (1 - B_{\text{hair}})
    \label{eq:cs}
\end{equation}
where $\Phi_{\text{strand}}(\cdot)$ and $\Phi_{\text{pose}}(\cdot)$ denote the feature extraction processes for strand map and human pose, respectively, both are neural networks in our implementation.
$\text{C}_{\text{t}}$ is the control signal for frame $\text{t}$ which has same resolution as the output video.
We show an example of control signal extraction in \Fig{fig:cs}.  
We obtain $\text{C}_{1:\text{T}}$ by applying this extraction to all  projected frames $\text{I}_{\text{proj},1:\text{T}}$, which serve as per-frame control signals for video diffusion model (\Sect{sec:method:diffusion}).

\subsection{Conditioned Video Diffusion for Hair Rendering}
\label{sec:method:diffusion}

With per-frame control signals $\text{C}_{1:\text{T}}$ and the reference image $\text{I}_\text{ref}$, we use a conditional video diffusion model $\mathcal{G}$ to generate the  video $\text{V}_{1:\text{T}}$ (Stage~3 in \Fig{fig:overview}):
\begin{equation}
\text{V}_{1:\text{T}} = \mathcal{G}\big(E_I(\text{I}_{\text{ref}}),\, E_{\text{C3D}}(\text{C}_{1:\text{T}}),  E_{\text{C2D}}(G(\text{I}_{\text{ref}})) \big)
\end{equation}
where $\text{C}_{1:\text{T}}$ provides frame-wise geometric constraints, while the reference image $\text{I}_{\text{ref}}$, encoded by $E_I$, specifies appearance constraints.
We use Wan~2.1 encoder~\cite{wan2025wan} as $E_I$.
Before feeding the control signals into the diffusion backbone, we encode them using a 3D convolutional network $E_{\text{C3D}}$. 
In addition, we extract geometric features (\Eqn{eq:cs}) from the reference image, encode them with a 2D convolutional network $E_{\text{C2D}}$.
Our design follows previous controllable diffusion models~\cite{zhang2023adding, wang2024unianimate, wang2025unianimate}.  
All encodings are fed into a Wan~2.1 diffusion transformer (DiT), which generates  video with controlled hair dynamics and human motion.

\textbf{Data Annotation Made Easy.}
No existing diffusion model supports our control signals. 
To address this, we fine-tune a pretrained model with paired control signals and RGB videos.
Through this training, the model learns to map control signals, together with a reference image, into video.
To obtain such data, we annotate RGB videos with control signals and sample one frame from each video as the reference image.
Specifically, we run geometric feature extraction (\Eqn{eq:cs}) on every video frame, with $B_{\text{hair}}$ estimated by a hair segmentation network. This gives us the control signals.
We show annotation  in \Fig{fig:annotation}.  

\proj makes training data annotation easier because inferring underlying physics from videos is challenging.
To enable high-quality training, we collect 30K real-world videos containing hair. 
We crop each video to focus the upper body and hair, and filter samples by resolution and aspect ratio. 
After annotation, we obtain a curated dataset of 10K videos.

\textbf{Training.}
To leverage priors from pretrained diffusion models, we initialize our model with pretrained weights from UniAnimate-DiT~\cite{wang2025unianimate} and fine-tune it with our annotated dataset to learn the mapping from new control signals to videos.  
During training, we keep the Wan encoder and decoder~\cite{wan2025wan} fixed.  
We fully fine-tune only the convolutional networks.  
For the DiT backbone, we apply LoRA~\cite{hu2021lora}, which adapts the model by adding only low-rank components to the weights. 
The motivation is that the DiT backbone contains most of the parameters and prior knowledge; using low-rank fine-tuning not only reduces computation but also prevents overfitting to our dataset.

\section{Evaluation}

\subsection{Experimental Setup}

\paragraph{Dataset.}  
We train \proj on a dataset (\Sect{sec:method:diffusion}) covering diverse videos. To characterize the training data, we annotate the visual appearance and motion of randomly sampled videos and report in the Supplementary Material \ref{sec:supp:dataset}.

For image animation evaluation, we use long-hair images from FFHQ-Wild~\cite{karras2019style} to assess rich hair dynamics, since short hair exhibits limited motion. 
For control-signal-to-video evaluation (\Sect{sec:eval:quant}), we generate $\sim$100 candidate videos with wind- or human-motion-driven hair dynamics using Google VEO~2~\cite{deepmind2025veo}, filter visually similar clips, and select 10 diverse videos as the evaluation set.

\paragraph{Implementation Details.}  
We describe the design choices of each module in \Fig{fig:overview}:
\begin{itemize}

\item In \textbf{Stage~1}, we adopt DiffLocks~\cite{rosu2025difflocks} to extract 3D hair strands from a single image, and use Blender's particle system~\cite{blender} to simulate hair dynamics. 
The simulator exposes configurable physical parameters, including hair properties such as mass, stiffness, and damping, and external forces such as wind, human motion, and gravity. We use these parameters as user controls for better controllability and keep unspecified parameters at their default values. Unless otherwise specified, we use mass = 0.1, stiffness = 6, damping = 9, wind strength = 10, and gravity = 1 (scale factor) as a stable default setting. 
For specific cases, users may tune these parameters, and future work could use vision-language models to predict suitable values.

\item 
In \textbf{Stage~2}, we use Blender’s camera system for perspective projection. 
The strand maps are extracted with a pretrained U-Net from HairStep~\cite{zheng2023hairstep},  
and 2D human poses are estimated using DWPose~\cite{yang2023effective}.  
\item 
In \textbf{Stage~3}, we follow the network design of UniAnimate-DiT~\cite{wang2025unianimate}
and adopt the encoder, decoder, and DiT from Wan~2.1~\cite{wan2025wan}.  
We set the output video resolution to $81 \times 832 \times 480$ (T$\times$H$\times$W) for all experiments. 
\end{itemize}

During training of diffusion model, we fix the encoder and decoder from Wan, fully finetune the convolutional network, 
and train the DiT with LoRA~\cite{hu2021lora} (rank = 128) to enable hair-conditioned video diffusion.  
Training is performed with a learning rate of $1\times 10^{-4}$, a batch size of 4, and 10K iterations, 
taking three days on four H100 80GB PCIe GPUs.

\paragraph{Runtime Analysis.}
{We report the runtime of \proj on a machine equipped with an AMD EPYC 9354 CPU and an NVIDIA RTX Pro 6000 Max-Q GPU for generating an 81-frame video.
Hair reconstruction takes 22 seconds, Blender simulation takes 326 seconds, control-signal extraction takes 21 seconds, and video diffusion takes 853 seconds.
The full pipeline therefore takes 1222 seconds, or about 20 minutes, per video.
The current system is designed for offline generation rather than real-time rendering.
The main runtime bottlenecks are Blender simulation, which lacks GPU acceleration in our setup, and high-quality video diffusion inference.
We treat real-time interactive inference as future work.}

\paragraph{Baselines.}  
Since no prior video generation method targets controllable hair physics, we compare \proj with two intuitive alternatives to show that \proj enables more precise and detailed control over hair dynamics:

\begin{itemize}
    \item \mode{Wan2.1}~\cite{wan2025wan}: a strong video diffusion model that supports image-to-video generation conditioned on text prompts. 
    We provide detailed text prompts to describe the  physics.
    \item \mode{UniAnimate-DiT}~\cite{wang2025unianimate}: a image-to-video generation framework conditioned on per-frame human pose and text.
\end{itemize}

\begin{figure*}[t]
  \centering
  \includegraphics[width=\textwidth]{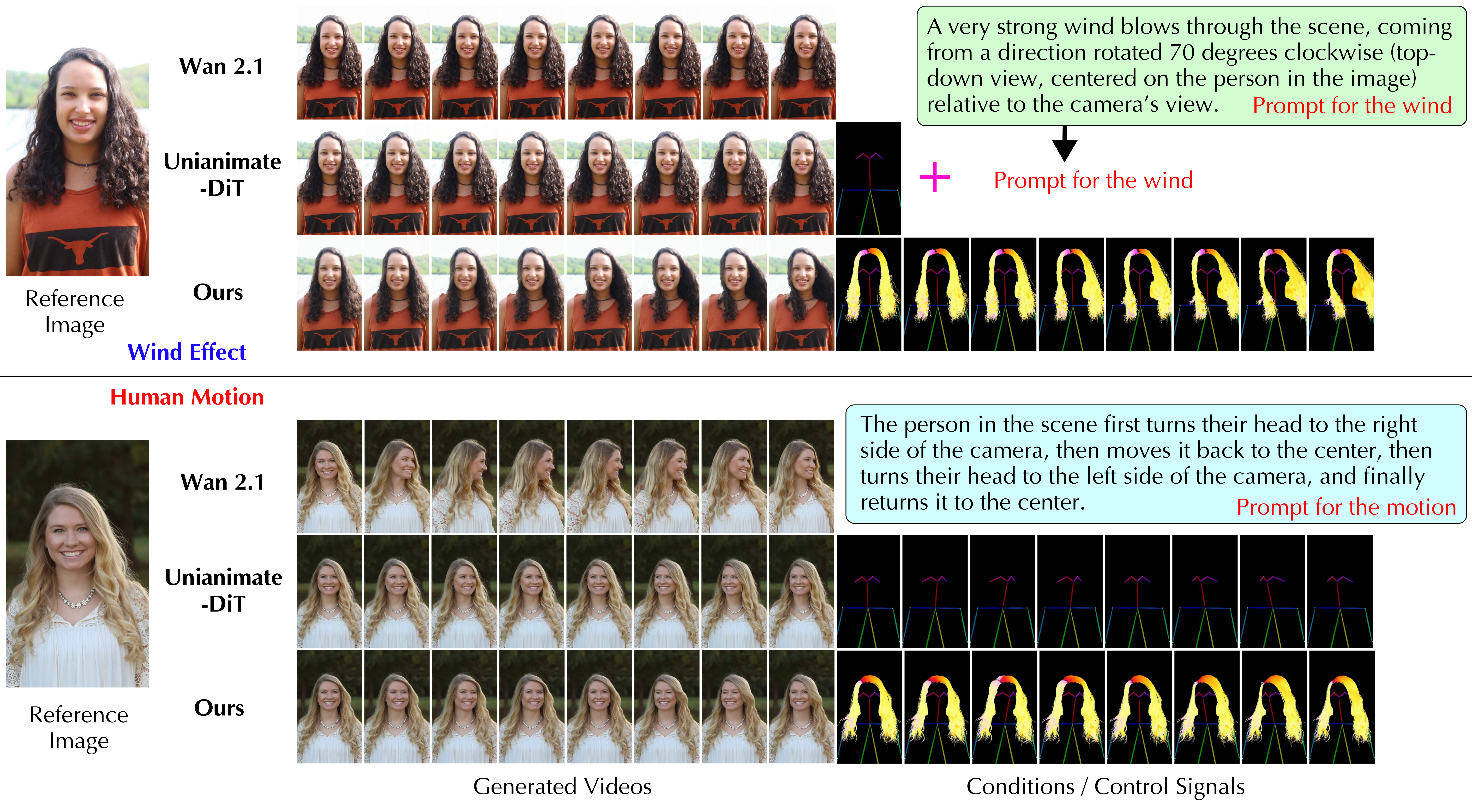} 
  \caption{Qualitative comparison with baselines under strong wind (top) and human motion (bottom). 
  Only \proj generates desired control of hair dynamics.}
  \label{fig:baselines}
\end{figure*}

\subsection{Qualitative Comparison}
\label{sec:eval:qual}
We evaluate \proj against baselines under two physics settings. 
Setting~1 applies strong wind to the hair, while Setting~2 applies head motion that induces hair dynamics
The results are shown in \Fig{fig:baselines}.

In Setting~1 (top), we add a strong wind to blow the hair of the reference image. 
The wind direction is rotated $70^\circ$ clockwise (top-down view) from the camera view, centered on the person.  
For the baselines, the only way to represent wind is through text prompts. 
We provided detailed prompts explicitly describing strong wind effects, as shown in \Fig{fig:baselines}. 
However, even with aggressive wording (e.g., ``strong wind''), both baselines fail to move the hair, highlighting the difficulty of controlling fine-grained physical effects via text conditions.  
In contrast, \mode{\proj} leverages control signals from a physics simulator and is able to drive hair motion in the expected direction.

In Setting~2 (bottom), the user’s head is required to first rotate to the right of the camera, then to the left, and finally return to the center.  
We evaluate two aspects: whether the head motion follows the specified sequence, and whether the hair dynamics are controllable.  
In \mode{Wan2.1}, we use detailed text prompt shown in \Fig{fig:baselines}.  
Although the prompt clearly specifies the intended motion, the generated video shows the head turning only to the right but never to the left, violating the required sequence.
\mode{UniAnimate-DiT} successfully controls the head pose, as it is conditioned on human poses, however, it cannot control hair dynamics and instead relies solely on the diffusion prior.  
In contrast, \mode{\proj} achieves accurate control of both head motion and hair dynamics by following simulator-provided signals, yielding more controllable results.

\subsection{Quantitative Comparison}
\label{sec:eval:quant}

\begin{table}[t]
\centering
\caption{Quantitative comparison between generated and ground-truth results.
}
\label{tbl:obj}
\footnotesize
\begin{tabular}{lccc}
\toprule
Method & PSNR $\uparrow$ & SSIM $\uparrow$ & LPIPS $\downarrow$ \\
\midrule
Wan~2.1~\cite{wan2025wan} & 11.65 & 0.518 & 0.499 \\
UniAnimate-DiT~\cite{wang2025unianimate} & 15.78 & 0.637 & 0.390 \\
\proj (Ours) & \textbf{17.15} & \textbf{0.668} & \textbf{0.370} \\
\bottomrule
\end{tabular}
\end{table}
We now evaluate qualitatively how well \proj leverages and follow the control signals. 
We use the control-signal-to-video reconstruction task, which is widely adopted in  controllable generation~\cite{wang2024unianimate, hu2024animate, karras2023dreampose}. 
Specifically, we treat videos generated by Google VEO~2~\cite{deepmind2025veo} as ground truth (the user-expected video) and extract control signals from them, which are then used to reconstruct the original videos.
For \mode{Wan2.1}, we use manually annotated text prompts as control signals.
For \mode{UniAnimate-DiT}, we include additional per-frame human poses extracted using DWPose~\cite{yang2023effective}.
For \mode{\proj}, control signals include per-frame human poses and hair strand maps, which is extracted using \Eqn{eq:cs}.
Our goal is to test if each  model can follow the control signals to reconstruct the expected video.

The results are shown in \Tbl{tbl:obj}, where we compare the generated and ground-truth videos using widely adopted frame-similarity metrics: PSNR~\cite{huynh2008scope}, SSIM~\cite{wang2004image}, and LPIPS~\cite{zhang2018unreasonable}. 
The results are averaged across all frames in all videos.
\mode{Wan 2.1} performs the worst, as text prompts cannot precisely capture per-frame physics.
\mode{UniAnimate-DiT} performs better since it has additional human poses input.
Among all methods, \mode{\proj} achieves the best results, achieving improvements of up to 47\%, 29\%, and 26\% in PSNR, SSIM, and LPIPS.  
This indicates: (1) \mode{\proj} has the highest physics controllability, and (2) \mode{\proj} can leverage hair-dynamics control signals (strand maps).

\subsection{Use Cases}
\label{sec:usecase}
To demonstrate the applicability of \proj, we showcase three representative use cases: dynamic hair try-on, bullet-time hair effects, and cinemagraphic. 

\textbf{Dynamic Hair Try-On.}  
Hair try-on is an important task: it allows users to preview whether a hairstyle suits them before making changes. 
However, prior work has focused only on static try-on~\cite{zhang2025stable, chung2025preserve}, limiting users’ understanding of dynamic hairstyle appearance.  
To enable dynamic hair try-on, we cascade \proj after the static hair try-on system HairFusion~\cite{chung2025preserve}.

\begin{figure*}[t]
  \centering
  \includegraphics[width=\textwidth]{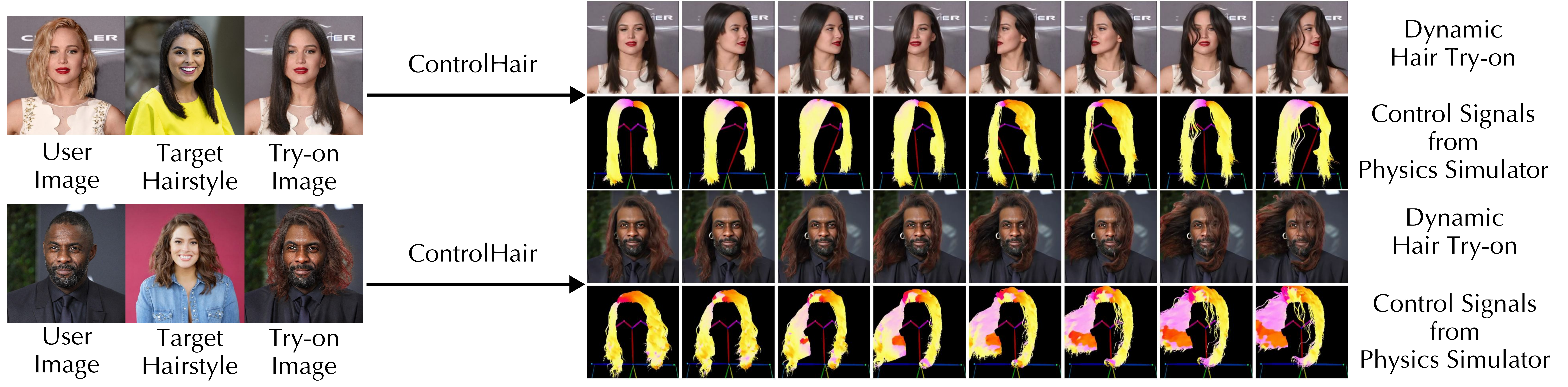} 
  \caption{Dynamic Hair Try-on using \proj. We cascade \proj after existing static image-based hair try-on frameworks to enable dynamic hair try-on.}
  \label{fig:tryon}
\end{figure*}

The results are shown in \Fig{fig:tryon}.  
Given a user image and a target hairstyle, we first apply HairFusion to obtain a static try-on image.  
We then use \proj to introduce dynamics such as head rotations (first row) or wind effects (second row), yielding the final dynamic hair try-on videos.  
The results show that by combining \proj with a static try-on system, we extend static hair try-on to dynamic hair try-on with detailed and controllable dynamics.

  \begin{figure*}[t]
    \centering
    \includegraphics[width=\textwidth]{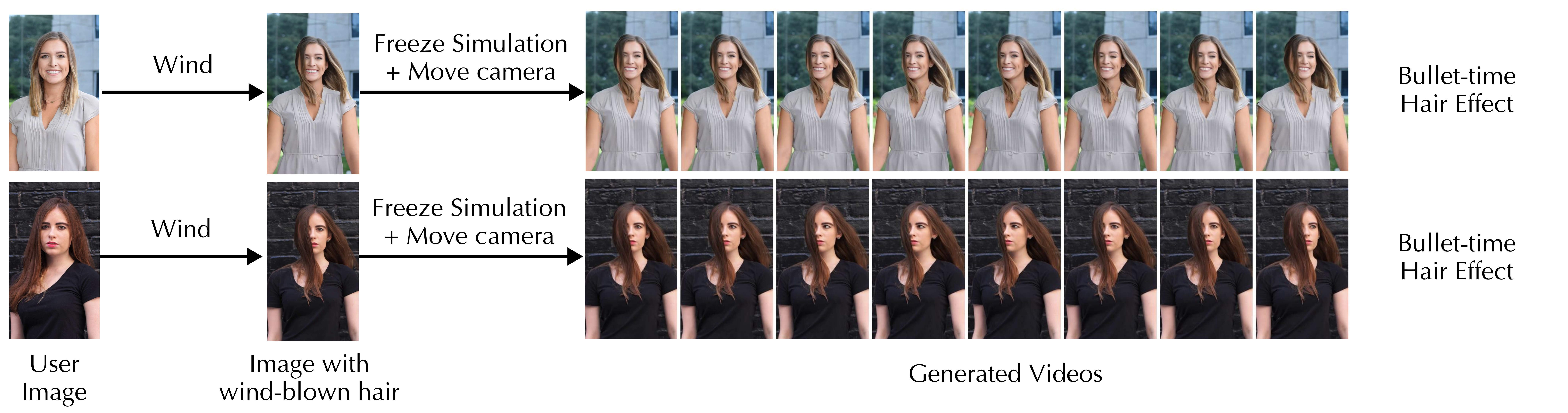} 
    \caption{Bullet-Time Effects. We freeze the physics simulator and rotate the camera around the user to produce the bullet-time effect.}
    \label{fig:bullet}
  \end{figure*}

  \begin{figure*}[t]
    \centering
    \includegraphics[width=\textwidth]{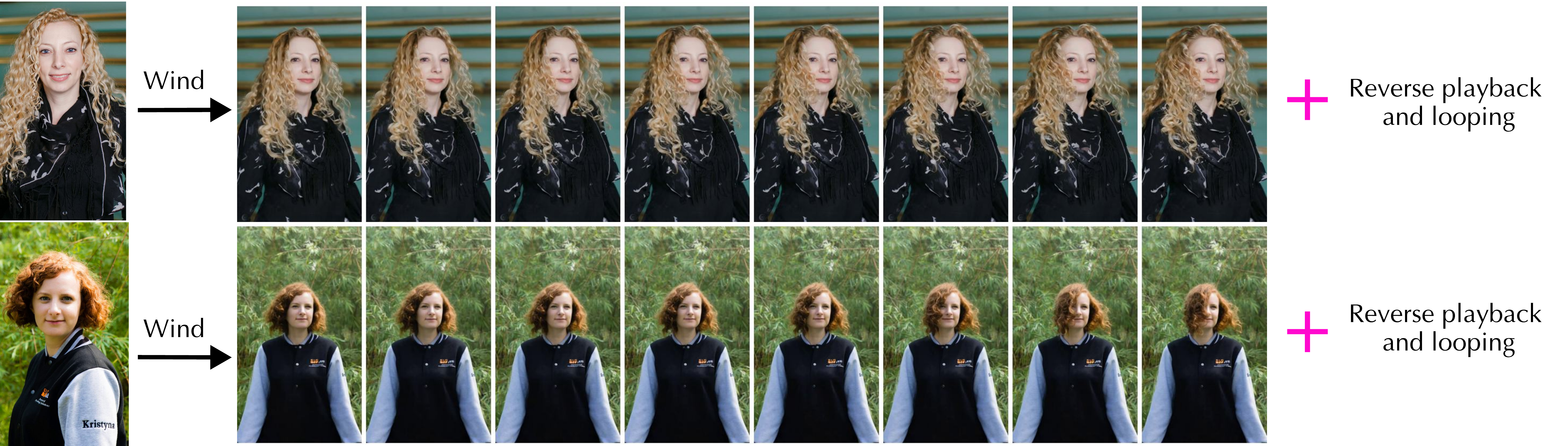} 
    \caption{Cinemagraphic. We animate only the hair and append reverse playback to the video, achieving a cinemagraphic effect.}
    \label{fig:cine}
  \end{figure*}

\textbf{Bullet-Time Hair Effects.}  
As discussed in \Sect{sec:method:ctrl_sig}, we can control the camera trajectory of video generation by manipulating the perspective projection. 
This enables bullet-time effect: at a specific moment of physics simulation, we can freeze the simulator to lock the hair geometry and render the scene from different viewpoints.
This  allows users to observe the hair from different perspectives at the same instant.
We showcase this application in \Fig{fig:bullet} by first applying a left-to-right wind in the physics simulator.  
We then freeze the simulator.  
After freezing, we rotate the camera around the person: $20^\circ$ to the right, followed by $40^\circ$ to the left.
The hair is frozen, and the camera trajectory is well-controlled.

\textbf{Cinemagraphic Effects.}  
Beyond animating images, \proj can generate unconventional physical effects such as zero gravity, extremely stiff or soft hair, or unusually strong wind.  
Such effects are difficult to generate using other diffusion models if they are absent from the training data.
To showcase this capability, we use \proj to generate videos with cinemagraphic effects.  
Specifically, we design the effect such that the human subject remains static while only the hair is animated, and the video is looping. 
Such effect is difficult to capture in the real world, and producing them requires post-processing.

To achieve this effect, we fix the human pose in the control signals while applying wind to the hair.  
Looping is enabled by appending a reversed playback segment to the video.  
The results are shown in \Fig{fig:cine}, where the model successfully freezes the human while animating the hair.  


\section{Discussion}
\label{sec:discussion}

\proj may also introduce two kinds of errors.
First, the 3D hair model estimated from a single input image in Stage~1 can be inaccurate because single-view reconstruction is under-constrained. The video diffusion model can sometimes correct these errors using the reference image, and we show such failure cases in the Supplementary Material~\ref{sec:supp:failure}. 
Future work could mitigate this issue by using multi-view priors~\cite{gao2024cat3d, shi2023mvdream, miao2024dsplats, tang2023emergent} to infer more accurate 3D hairstyles.

Second, the performance of the physics simulator affects the final results. If the chosen simulator is imperfect, the generated video may look unrealistic.
However, we observe that when the physics simulator is sufficiently accurate, as in the control-signal-to-video tasks (\Sect{sec:eval:quant}), our diffusion model can generate highly realistic results.
To further improve simulation quality, one can adopt a more advanced simulator~\cite{sidefx2025houdini} or even a neural simulator learned from  data~\cite{lin2025neuralocks}.

\textbf{Extreme Cases.} We also tested extreme scenarios such as multiple wind sources combined with drastic human motion.
Experimental results demonstrate that \proj can handle these extreme cases.
We found Blender physics simulator might diverge under certain extreme physics or very short hair; however, this is related to the choice of simulator and is independent of the our framework.
We also show full-body and out-of-distribution examples with large motion in the Supplementary Material~\ref{sec:supp:generalization}; large motion may still cause blurred frames, a common limitation of current video diffusion models.

\section{Conclusion}
We presented \proj, a three-stage pipeline decouples physics reasoning from video generation, integrates physics simulators with video diffusion to enable fine-grained control over the physics.
Trained on a curated dataset, \proj outperforms  baselines in both qualitative and quantitative evaluations. 
\proj also enables diverse applications, including dynamic hairstyle try-on, bullet-time effects, and cinemagraphic.
The hybrid design in \proj opens possibilities for embedding physics into generative video models, paving the way toward controllable and physically grounded rendering.

\bibliographystyle{splncs04}
\bibliography{hair}

\clearpage
\appendix
\renewcommand{\thesection}{\Alph{section}}

\section{Supplementary Material}
\label{sec:supp:overview}
This supplementary material provides additional details and results for the main paper.
This PDF focuses on textual discussions and results suitably presented as images.
Since our work generates videos, most experimental results are best viewed in motion.
We encourage readers to visit our project website, \href{https://linwk20.github.io/controlhair-web/}{https://linwk20.github.io/controlhair-web/}, to view the video results.

\section{Dataset Analysis}
\label{sec:supp:dataset}

To verify the diversity of our training dataset, we scrutinized hundreds of randomly sampled videos, confirming that the dataset includes people from different regions, genders, and ethnic groups, with diverse hair lengths, styles, and textures.
Concretely, we analyzed 100 randomly sampled videos and annotated the following attributes: gender (Female: 71\%, Male: 29\%), hair length (above shoulder: 62\% vs.\ below shoulder: 38\%), and style (straight: 64\% vs.\ curly: 36\%).

{We further annotate for visual appearance(skin tone and hair color) and motion strength.
For visual appearance, we use coarse observable labels from video frames only. 
For skin tone, 72\% of samples are labeled Light and 28\% are labeled Medium and Dark.
For hair color, the samples contain black/dark brown hair (40\%), blond/light hair (23\%), brown hair (23\%), gray/white hair (7\%), red/dyed hair (4\%), and cases where hair is not visible (3\%).
For motion strength, 53\% are labeled large-motion and 47\% are labeled small-motion; this label jointly considers hair motion, human/body motion, and camera motion. }

Since our video diffusion model is trained on large-scale videos, it can handle common hair styles, including kinky hair.
However, our physics simulator (Blender) and the strand-estimation model (DiffLocks~\cite{rosu2025difflocks}) currently cannot support kinky hair well.
Supporting kinky hair would require a different reconstruction method and a simulator tailored for highly curled hair; 
As long as the appropriate modules are used, we can support kinky hair dynamics through on-site swapping and plug-and-play.
We leave this as future work.

\section{Full-Body and Extreme Motion Results}
\label{sec:supp:generalization}

We further evaluate \proj on challenging inputs beyond the portrait-style examples in the main paper, including full-body subjects and rapid hair/body motion using the control-signal-to-video protocol (Sec. 5.3). 

\begin{figure*}[h]
  \centering
  \includegraphics[width=\textwidth]{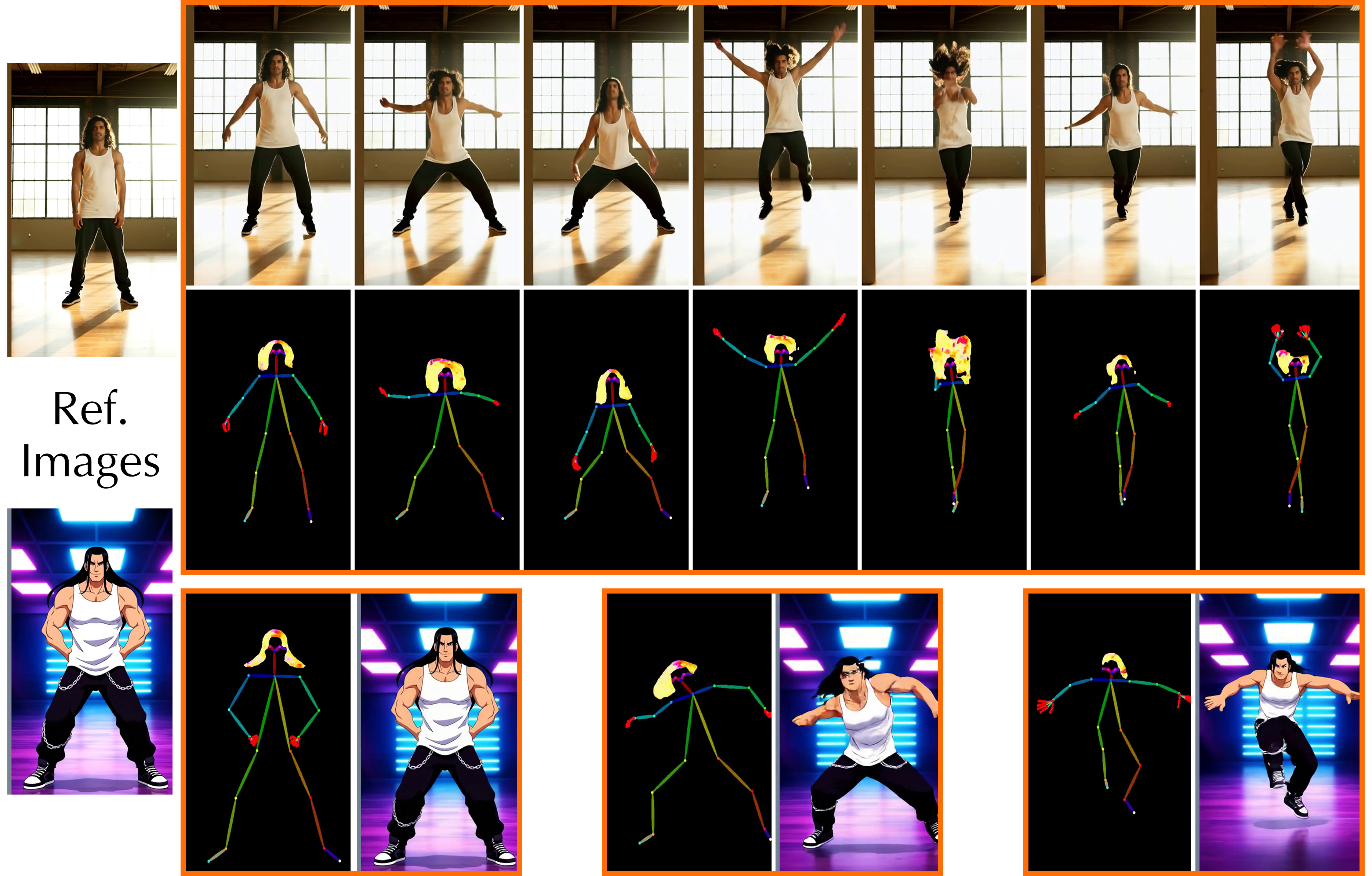}
  \caption{Generalization results on full-body and anime-style out-of-domain inputs with rapid hair/body motion.}
  \label{fig:supp:generalize}
\end{figure*}

As shown in \Fig{fig:supp:generalize}, \proj can follow large human motions while preserving the reference appearance and generating dynamic hair responses from the control signals. 
We also test an anime-style out-of-domain input with strong body motion, where the generated frames still follow the provided pose and hair controls. 
These examples suggest that \proj can generalize to broader human-centric inputs, while our target scope remains portrait and digital-human hair rendering rather than unconstrained open-world generation.

\section{Failure Case and Error Propagation}
\label{sec:supp:failure}
\label{sec:supp:autocorrect}

\begin{figure*}[t]
  \centering
  \includegraphics[width=\textwidth]{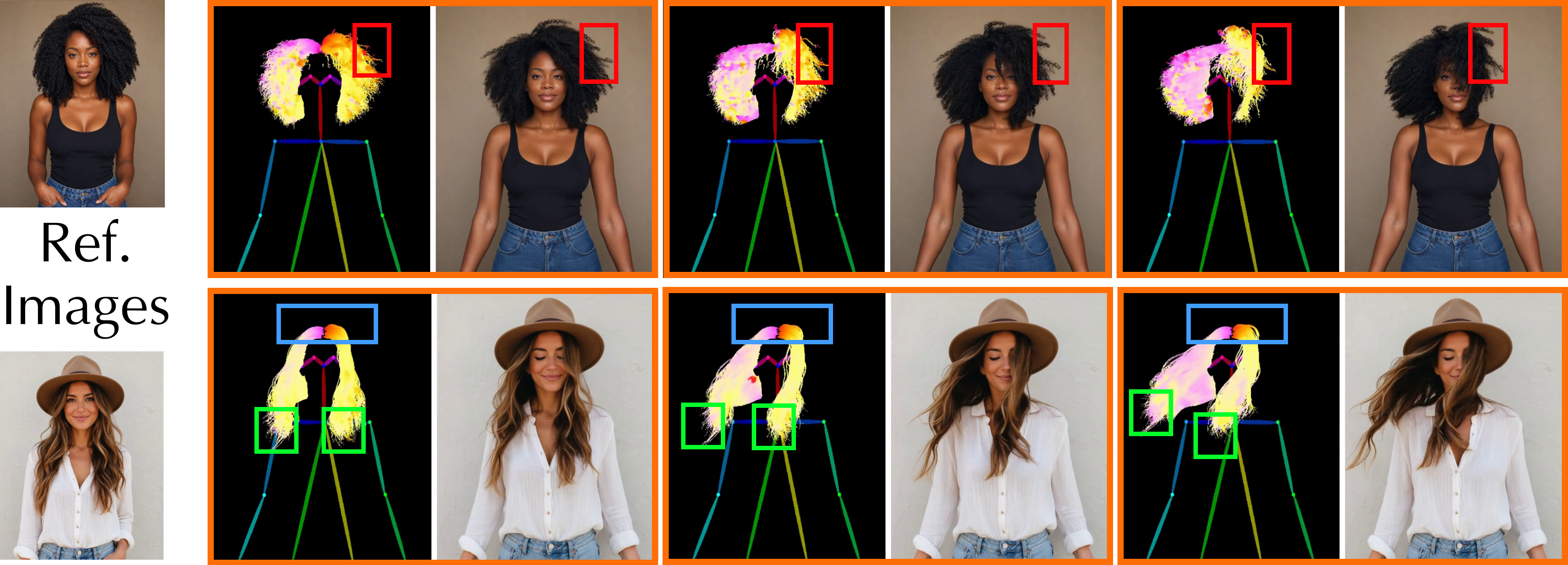}
  \caption{Failure cases and error propagation for kinky hair (first row) and hair accessories (second row).}
  \label{fig:supp:failure}
\end{figure*}

 As discussed in the main paper (\Sect{sec:discussion}), upstream reconstruction and simulation errors can affect the final rendering. \Fig{fig:supp:failure} shows two representative failure cases. 
 
\begin{figure*}[t]
  \centering
  \includegraphics[width=\textwidth]{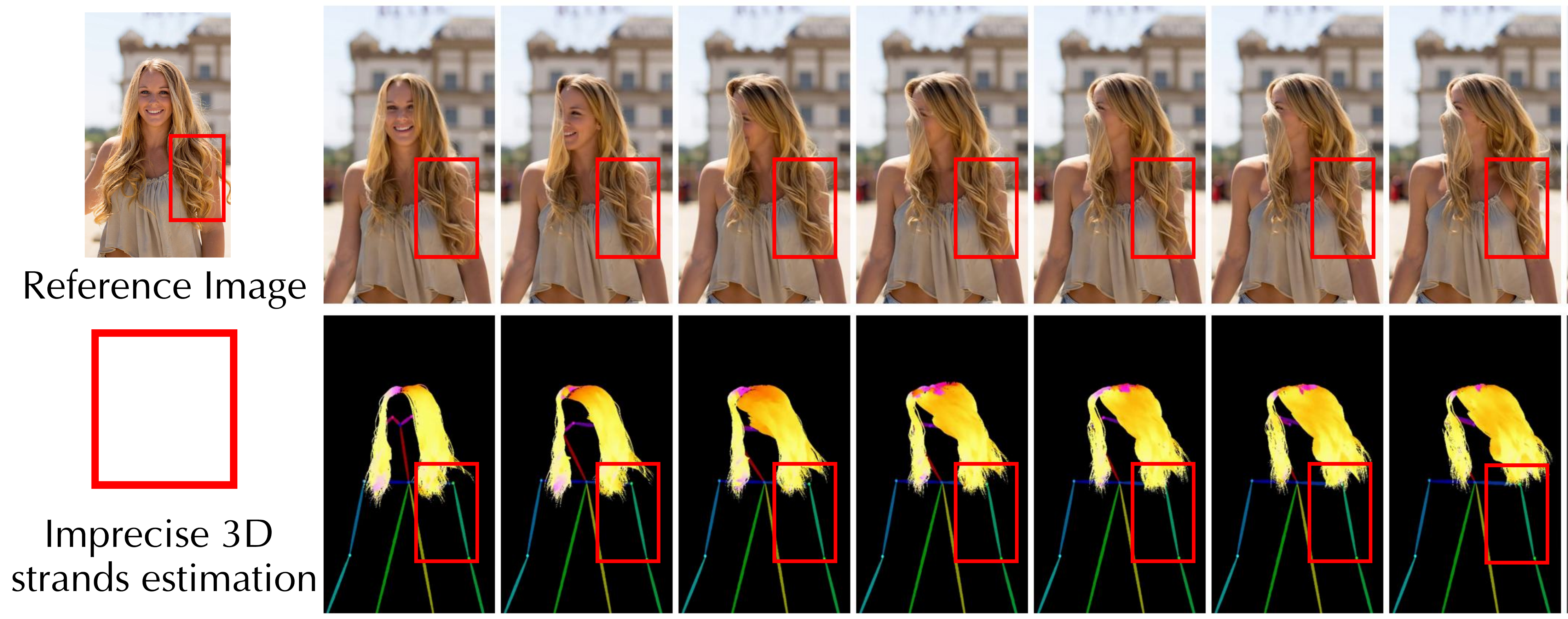}
  \caption{Auto correction. 3D hair strands estimation from a single image may introduce errors, but the diffusion model often corrects moderate errors by leveraging the reference image.}
  \label{fig:autocorrect}
\end{figure*}

 For kinky hair, the current strand reconstruction and Blender simulation may produce inaccurate control signals with protruding strands, and such errors can propagate to the generated video. 
 For hair accessories and long hair, the reconstruction stage may fail to capture hat geometry and may produce inaccurate hair length for long hair (also observed in \Fig{fig:autocorrect}). 
 These cases motivate future work on hairstyle-aware reconstruction, accessory-aware geometry estimation, and stronger simulators.


{\textbf{Error correction and propagation.} The video diffusion model can also correct moderate upstream errors when the reference image provides strong appearance evidence. \Fig{fig:autocorrect} and second row in \Fig{fig:supp:failure} show examples where imperfect hair strand estimation produces controls with shorter hair. 
The diffusion model leverages the reference image to adjust the generated hair length, so the final video exhibits no noticeable hairstyle change. 
The hat example in \Fig{fig:supp:failure} similarly indicates that the video model can restore missing appearance details and compensate for moderate control errors, while severe geometry errors may still propagate.}

\end{document}